\documentclass[12pt,preprint]{aastex}
\slugcomment{}

\newcommand{\beq}[1]{\begin{equation}\label{#1}}
 \newcommand{\eeq}{\end{equation}}
 \newcommand{\bea}{\begin{eqnarray}}
 \newcommand{\eea}{\end{eqnarray}}

\begin{document}

\title{Source plane reconstruction of the giant gravitational arc in Abell~2667:
a candidate Wolf-Rayet galaxy at $z\sim1$}

\author{Shuo Cao$^{1,2}$, Giovanni Covone$^{2,3}$, Eric Jullo$^{4}$, Johan Richard$^5$, Luca Izzo$^{6,7}$, and Zong-Hong Zhu$^{1\ast}$}

\affil{$^1$ Department of Astronomy, Beijing Normal University, Beijing 100875, China; zhuzh@bnu.edu.cn \\
$^2$ Dipartimento di Scienze Fisiche, Universit\`a di Napoli ''Federico II'', Via Cinthia, I-80126 Napoli, Italy \\
$^3$ INFN Sez. di Napoli, Compl. Univ. Monte S. Angelo, Via Cinthia, I-80126 Napoli, Italy \\
$^4$ OAMP, Laboratoire d'Astrophysique de Marseille, UMR6110, traverse du Siphon, 13012 Marseille, France \\
$^5$ CRAL, Observatoire de Lyon, Universit¡äe Lyon 1, 9 Avenue Ch. Andr¡äe, 69561 Saint Genis Laval Cedex, France \\
$^6$ Dip. di Fisica, Sapienza Universit`a di Roma, Piazzale Aldo Moro 5, I-00185 Rome, Italy \\
$^7$ ICRANet, Piazza della Repubblica 10, I-65122 Pescara, Italy}

\begin{abstract}
We present a new analysis of HST, Spitzer telescope imaging and VLT imaging and spectroscopic data
of a bright lensed galaxy at $z$=1.0334 in the lensing cluster Abell~2667. Using this high-resolution imaging we present an updated lens model that allows us to fully understand the lensing geometry and reconstruct the lensed galaxy in the source plane. This giant arc gives a unique opportunity to peer
into the structure of a high-redshift disk galaxy. We find that the lensed galaxy of Abell 2667 is a typical spiral galaxy with morphology similar to the structure of its counterparts at higher redshift $z\sim 2$. The surface brightness of the reconstructed source galaxy in the z$_{850}$ band reveals the central surface brightness $I(0)=20.28\pm0.22$ mag arcsec$^{-2}$ and the characteristic radius $r_s=2.01\pm0.16$ kpc at redshift $z \sim 1$.
The morphological reconstruction in different bands shows obvious negative radial color gradients
for this galaxy. Moreover, the redder central bulge tends to contain a metal-rich stellar population, rather than being heavily reddened by dust due to high and patchy obscuration. We analyze the VIMOS/IFU spectroscopic data and find that, in the given wavelength range ($\sim 1800-3200$ \AA), the combined arc spectrum of the source galaxy is characterized by a strong continuum emission with strong UV absorption lines (FeII and MgII) and shows the features of a typical starburst Wolf-Rayet galaxy NGC5253. More specifically, we have measured the EWs of FeII and MgII lines in the Abell 2667 spectrum, and obtained similar values for the same wavelength interval of the NGC5253 spectrum. Marginal evidence for CIII] 1909 emission at the edge of the grism range further confirms our expectation.

\end{abstract}
\keywords{Galaxies: clusters: individual: Abell 2667 --- Gravitational lensing: strong}

\section{Introduction}

Considerable advances have been made over the past 15
years to chart and investigate the properties and evolution
of the disk galaxies, which may provide an insight into the
galaxy formation process in a dark matter-dominated universe.
Moreover, it is well known that the galaxy stellar disks
can not only be thickened by interactions with low-mass halos \citep{Benson04}, but also be entirely disrupted by the
galaxies or halos with a few of the galaxy masses \citep{Koda08}. Therefore, the environments where these galaxies are
located also play a significant role in the basic observational
features of the disk galaxy population.

HST imaging and spectroscopy of high-redshift disk galaxies
provide critical diagnostics of their formation and evolution
(e.g., \citet{Law09,Cassata10}). However, there are several observational challenges in pursuing the evolution history of disk galaxies beyond $z\simeq 1$. On the one hand, observations of low surface brightness galaxies at $z\simeq 1$ are very challenging, thus limiting the size of the available data; on the other hand, at high redshift
it becomes increasingly more difficult to resolve the spatial structure within galaxies, greatly restraining the possibility to understand the exact position where processes (like star formation) occur. Consequently, gravitationally lensing galaxy clusters provide a unique and detailed view of
the high-redshift universe, allowing to probe the faint population of distant, background galaxies. This technique, often
referred to as ¡±gravitational telescope¡±, was first proposed by Zwicky (1937) and has developed into an important astrophysical
tool since the advent of the Hubble Space Telescope, since lensed galaxies are magnified by typically 1-3 magnitudes
and spatial sizes are correspondingly stretched \citep{Kneib04,Bradley08,Zheng09}.

A magnificent gravitational lensing example is provided by the galaxy cluster Abell 370 at $z=0.375$ \citep{Soucail87,Lynds89} showing a giant gravitational arc ($z=0.725$) \citep{Richard10} that has been extensively analyzed with the multiband HST/ACS images. The galaxy cluster has a large Einstein radius, $\theta_E=39\pm2"$, for a source redshift
at $z=2$ and strong magnification in the central region make Abell 370 one of the best clusters to search for high redshift
galaxies. The first metallicity gradient measurement for a face-on spiral galaxy at $z\sim 1.5$ was also discussed by \citet{Yuan11}, through the analysis of the arc produced by the massive galaxy cluster MACS J1149.5+2223 at $z=0.544$. More recently, \citet{Sharon12} successfully reconstructed
the source plane of a bright lensed galaxy at $z=1.7$ magnified by the lensing cluster RCS2 032727-132623, by using
a new high-resolution imaging from Hubble Space Telescope/Wide Field Camera3 (HST/WFC3). A similar rare case was discovered in VIMOS Integral Field Unit (IFU) spectroscopic observations of the X-ray luminous cluster Abell 2667 ($z=0.233$).

The galaxy cluster Abell 2667 is another remarkable example of ¡±gravitational telescope¡±. It is among the most
luminous galaxy clusters in the X-ray sky \citep{Ebeling96}, with a regular X-ray morphology \citep{Newman13,Weissmann13}
and a cool core \citep{Covone06a}. Redshift survey in the central acrmin$^2$ reveals a unimodal velocity distribution of the member galaxies \citep{Covone06b}. Abell 2667 shows a remarkable giant gravitational arc \citep{Sand05} originated from a blue star-forming galaxy at $z=1.034 \pm 0.005$ \citep{Covone06b}, and a system of lensed sources at high redshift \citep{Laporte11,Laporte12}. \citet{Covone06b} performed a wide field integral field spectroscopy survey of the central arc min square, using the VIMOS Integral Field Unit (IFU) and built a strong lensing model using HST/WFPC2 data. \citet{Yuan12} have used this lensing model to derive the source-plane morphology from the archive WFPC2 data. They analyzed Keck II/ OSIRIS spatially resolved NIR spectra
and found that the higher [NII]/H$\alpha$ ratio in the outer region might be contaminated by a significant fraction of shock excitation due to galactic outflows.

In this paper, we extend the previous analysis on the magnified source of the giant arc in Abell 2667, based on HST/ACS $z_{850}$ imaging data and a set of complementary imaging and spectroscopic data. We improve the previous lensing model, produce a new inversion of the source galaxy and study in detail the color gradient in this $\sim 1$ system. The paper is organized as follows. In Section~\ref{s.observations}, we present the used observational data used in this work. The lensing analysis including a new lens model is presented in Section~\ref{s.lensing}. Section~\ref{s.property} shows the physical conditions and main properties of the reconstructed source galaxy. Discussion and conclusions appear in Section~\ref{s.conclusions}. We assume a flat cosmology with $\Omega_{\Lambda} = 0.7$, $\Omega_{m} =0.3$, and $H_0 = 70$ km s$^{-1}$ Mpc$^{-1}$. All magnitudes refer to those in the AB system.

\section{Observations}\label{s.observations}

In this section we will present the imaging and spectroscopic data used in the analysis of Abell 2667. We use the imaging data from the Hubble Space Telescope, taken with the Advanced Camera for Surveys (ACS; Program 10504, PI R. Ellis; filter F850LP), Wide-Field Planetary Camera 2 (WFPC2; Program 8882, PI Allen; filters F450W, F606W, F814W), and Near Infrared Camera and
Multi-Object Spectrometer (NICMOS; Program 10504, PI R. Ellis) and ground-based data from ISAAC at VLT (Program
71.A-0428, PI J.-P. Kneib) through the F110W, F160W filters. HST high-resolution images are essential to build an accurate model of the gravitational lens,
while the large multi-wavelength coverage allows us to constrain the properties, for example, the color gradient of the lensed source in different bands.

The HST/WFPC2 observations were obtained on October 10-11, 2001, for a total of 12000/5000/5000s in the F450W/F606W/F814W bands, respectively. Final data have the depth of $\simeq 25$ mag. Compared to the WFPC2 observations, the HST/ACS data in the F850LP band \textbf{cover} a square
field of about 18 acrmin$^2$ are deeper and have higher resolution. Final data were obtained on July 21, 2006 for a total of 8765s, which have the pixel scale of $0.05\arcsec$, with the depth of $z_{850} \simeq 28$. The near infrared imaging data are observed from NICMOS in broad, medium, and narrow band
filters, broad-band imaging polarimetry, coronographic imaging,
and slitless grism spectroscopy. The NICMOS observations were obtained on August 6, 2006, for a total of 4608/7104s in the F110W/F160W
bands with the depth of $\simeq 27$ mag. Moreover, the mid-infrared data for Abell 2667, which were obtained with Spitzer/IRAC (Program 83, PI G. Rieke; and Program 60034, PI E. Egami), will also be used for multi-band photometric measurements presented in Section~\ref{s.property}. The final Spitzer/IRAC data were acquired using four broad-band filters in the IR channel, and centered at 3.6, 4.5, 5.8 and 8.0 microns.

The central region of the galaxy cluster was observed with the VIMOS Integral Field Unit \citep{Covone06b,Grin07} mounted on VLT
Melipal, with a f.o.v. of $54'' \times 54''$, fiber size $0.66''$, and the low-resolution red grism ($R\sim 250$), covering
from about 3900 to 5300 {\AA}. \citet{Covone06b} extracted 22 galaxy spectra and the spatially resolved spectra of three lensed images of the gravitational
arc.

\begin{figure*}
\begin{center}
\includegraphics[width=13cm]{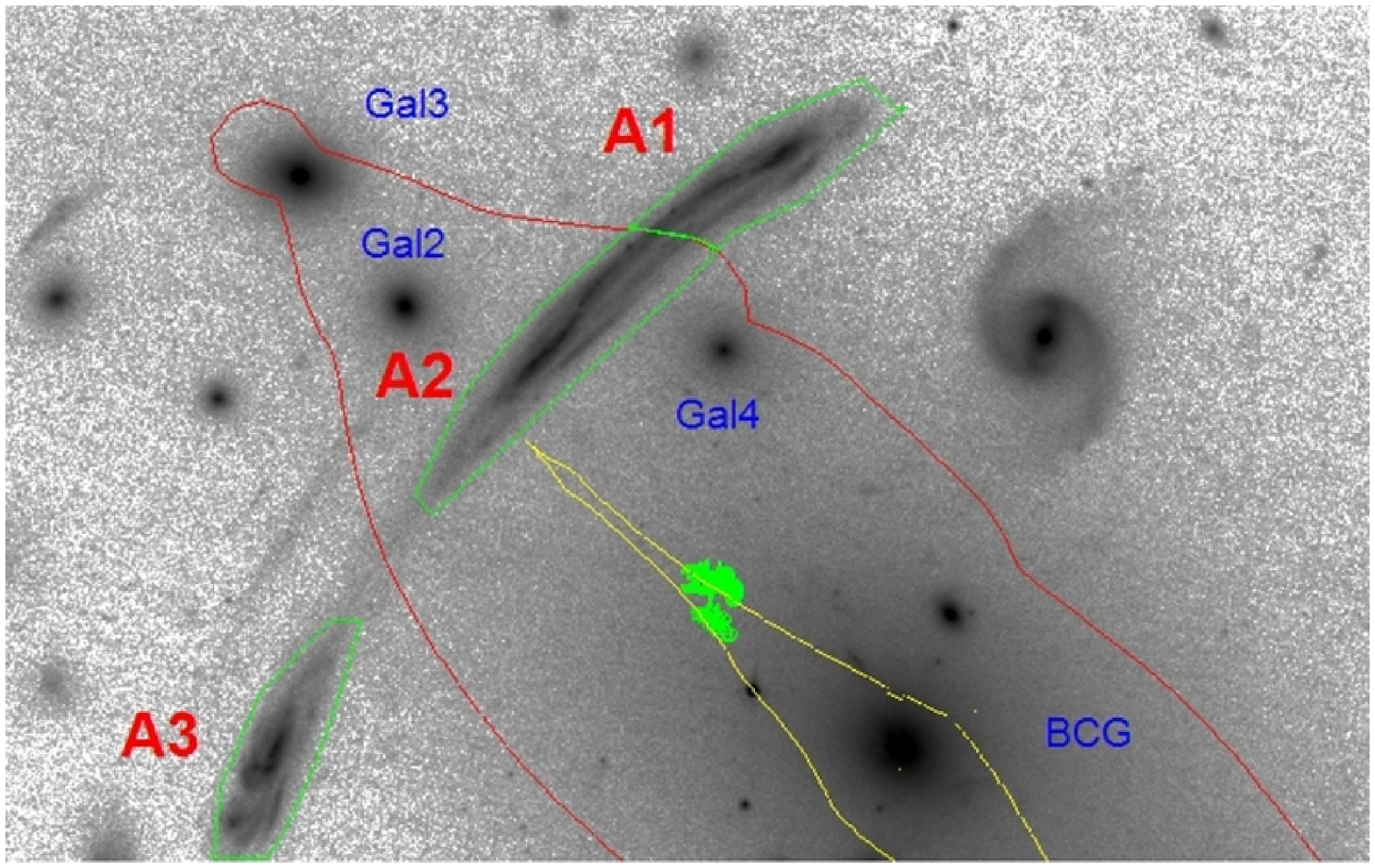}
\includegraphics[width=6.3cm]{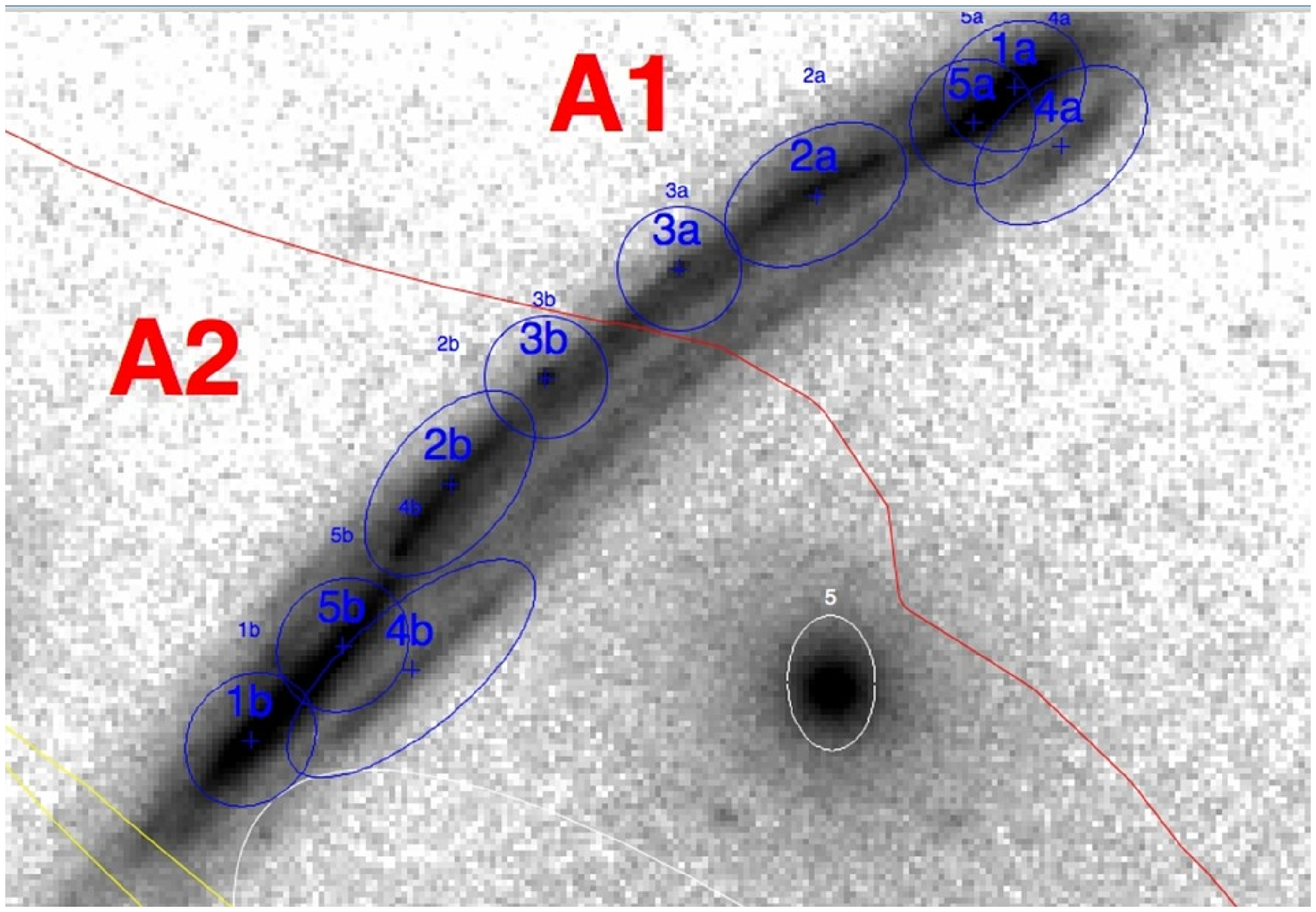} \includegraphics[width=6.7cm]{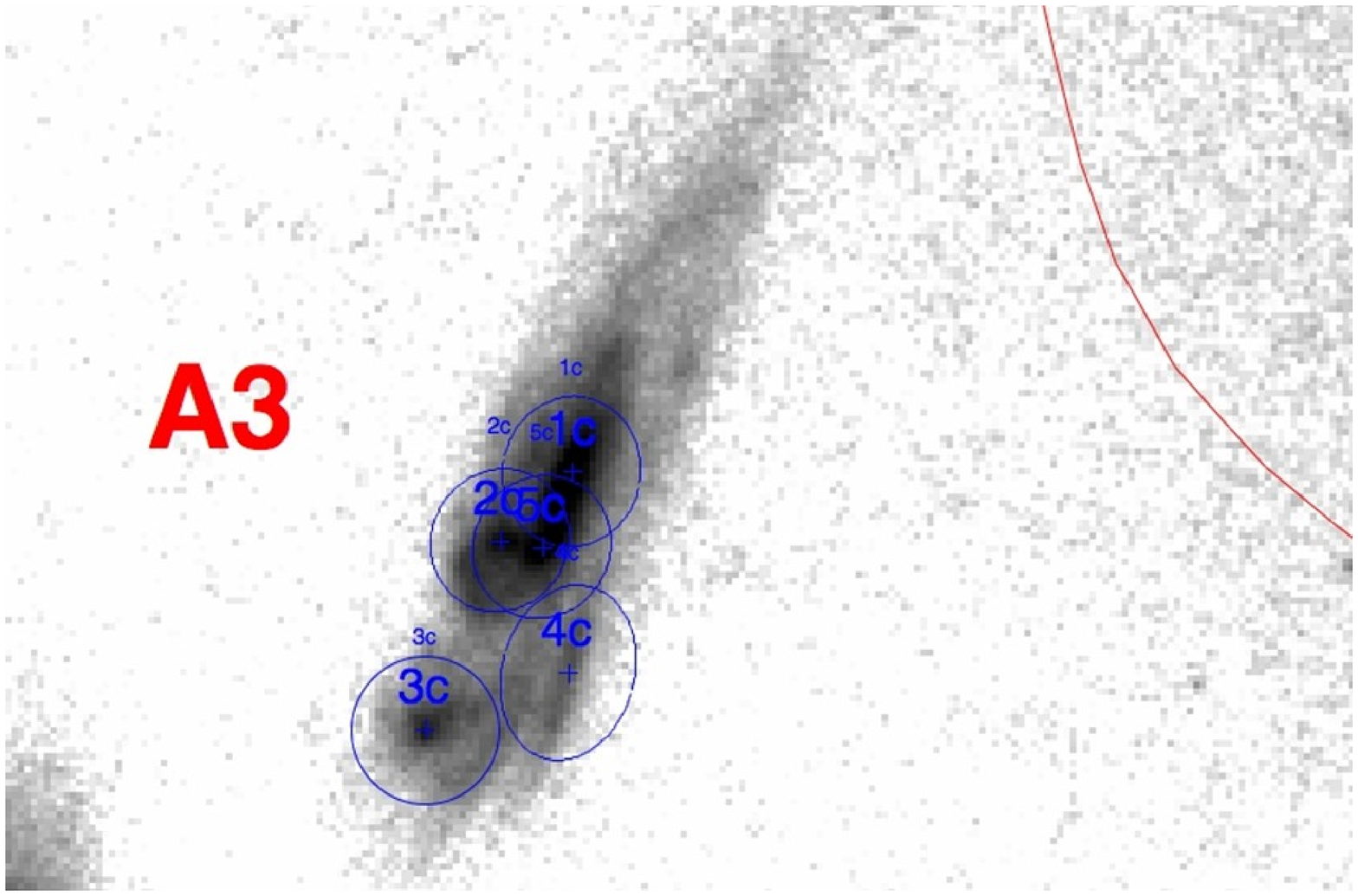}
\caption{ HST/ACS image of Abell 2667 in the $z_{850}$ band. (Top panel) Constraint results with the image-plane critical curves overplotted
in red and source-plane caustic in yellow. The favorite source positions are denoted
by green ellipses. The giant arc is indicated as A1-A3 within three green polygonal apertures. North is to the top and East to the left. (Bottom panels): location of matched regions identified with colored ellipses in each part of the giant arc.
\label{fig.constraint}}
\end{center}
\end{figure*}

\section{Strong Lensing Mass Model}\label{s.lensing}

We used the publicly-available software LENSTOOL \footnote{http://projets.lam.fr/projects/lenstool/wiki} \citep{Jullo07} to compute the best-fit lens model. LENSTOOL makes use of a Monte Carlo Markov Chain (MCMC) optimization method.

The source galaxy is multiply-imaged into three images
(A1, A2, A3; see Fig.~\ref{fig.constraint})\footnote{Hereafter, we use the same notation for the lensed images as in Covone
et al. (2006b).}: two of them (A1, A2) are merged,
forming the giant gravitational arc located $14\arcsec$ North-East of
the brightest cluster galaxy (BCG), while A3 constitutes an
isolated system. The clear mirror symmetry along the giant
arc provides additional constraints on the location of the critical
line at the arc redshift. HST/ACS data allow to identify
unambiguously multiply-imaged subregions within the three
images (Fig.~\ref{fig.constraint}b,c).

We use the HST/ACS to build a new lensing model of
the matter distribution in the cluster core. The positions of
the images are the main input observational data to constrain
the lens model parameters. We determine the position
of the brightest peak for each multiply lensed
feature in the arc, assuming a position error of $0.05\arcsec$ (the
spatial resolution of the HST/ACS data per pixel), see
also \citet{Faure11,Cao13,Collett14}.

To model the mass distribution, we include both a cluster mass-scale component (representing the contribution of the dark matter and baryonic gas halo)
and cluster galaxy mass components \citep{Kneib96,Smith05}.
Bright cluster galaxies within the central $\sim1 \, {\rm arcmin}^2$
belonging to the galaxy cluster are selected according to their redshift,
while massive galaxies along the line of sight are also included,
with their lensing properties rescaled at the cluster redshift. All the model components (including the large scale cluster halo, the BCG, and individual galaxies) are parameterized using a pseudo-isothermal mass distribution model (PIEMD, \citet{Kassiola93}), which is determined by its position $x$, $y$; ellipticity $e=(a^2-b^2)/(a^2+b^2)$, where $a$ and $b$ are the semi major and semi minor axes; a position angle $\theta$;
the velocity dispersion $\sigma_0$; a core radius $r_{core}$; and a truncation radius $r_{cut}$.

The best-fit lensing mass model is found by fitting the position,
ellipticity, orientation and mass parameters (velocity dispersion,
core and truncation radii) of the cluster-scale component.
The center of the BCG (x and y) is fixed at the observed
values in the ACS image, while the other parameters (the ellipticity,
orientation, and mass parameters) are set free. Other
galaxies are individually chosen to have the same position, ellipticity
and orientation as of their corresponding $z_{850}$-band
image. The effect of the mass distributions of the most massive
member galaxies Gal2-Gal4 are taken into consideration
with only one free parameter, the central velocity dispersion, $\sigma_0$. By assuming a \citet{Faber76} relation and a
global mass-to-light ratio (M/L), the mass of the other galaxies
are scaled according to their K-band luminosity estimated
by using a typical E/S0 spectral energy distribution \citep{Covone06b}. Therefore, the truncation radius and the velocity
dispersion are the two independent parameters of the
ensemble of cluster galaxies. The best-fit values of the model
parameters and their uncertainties are summarized in Table~\ref{t.parameters}.
The mean source-plane rms is $\simeq0.10 \arcsec$.
Concerning the ratio of the sizes between the image-plane
and the source-plane reconstruction, we obtain the total flux
magnification for the giant arc $\mu=14.0\pm 2.0$, consistent with
the result of \citet{Yuan12}.

\begin{table*}
\caption{ Best-fit lens model parameters: all coordinates are measured relative to the center of the BCG, at [RA, Dec]=[357.914250, -26.084105]. \label{t.parameters}}
\begin{center}{\scriptsize%\footnotesize
\begin{tabular}{lccccccc}

\hline
 Halo    & RA           & Dec         & $e$  &$\theta$ &$r_{\rm core}$ &$r_{\rm cut}$ &$\sigma_0$ \\
 (PIEMD) & ($\arcsec$)  & ($\arcsec$) &    &(deg)    &($\arcsec$)          &($\arcsec$)         &(km $s^{-1}$) \\
\hline
 Cluster  & $-0.15\pm1.54$   & $-0.54\pm1.47$  & $0.58\pm0.15$  & $-42.43\pm3.43$   & $15.32\pm4.97$  & $228.30\pm140.08$ & $887\pm116$   \\
 BCG      & 0     & 0     & $0.21\pm0.20$  & $-43.52\pm16.89$ & $0.85\pm0.28$ &$44.30\pm13.48$ & $169\pm43$ \\
 Gal2     & -12.71 & 11.34 &  0.25 & -41.40 & \nodata & \nodata & $65\pm20$ \\
 Gal3     & -15.42 & 14.66 &  0.17 & -9.40  & \nodata & \nodata & $123\pm28$ \\
 Gal4     &  -4.53 & 10.21 &  0.41 & -88.40  & \nodata & \nodata & $103\pm25$ \\
 L* galaxy  & \nodata & \nodata & \nodata & \nodata & \nodata  &   $22.03\pm6.83$&  $111\pm28$ \\
\hline

\end{tabular}}
\end{center}
\end{table*}

\section{Properties of the source galaxy}\label{s.property}

\subsection{Source reconstruction}\label{s.source}

The gravitational lensing model enables us to derive the
source plane properties independently through each of the
three images A1, A2, A3. The source plane reconstruction
is performed by using the task {\tt cleanlens} in \texttt{LENSTOOL},
computing the corresponding position in the source plane
for each point of the image plane through the best-fit lens
model. There are two related subsampling parameters in
our analysis, {\it echant} and {\it s-echant}. With $\it echant$=2, each pixel in the CCD frame will be cut into 4 smaller pixels
during the calculation of the the source frame. The {\it s-echant}
parameter plays the same role when subsampling
the frame in the source plane. In our analysis, {\it echant} and {\it s-echant} are both fixed at 2.

\begin{figure*}
\begin{center}
\includegraphics[width=6cm]{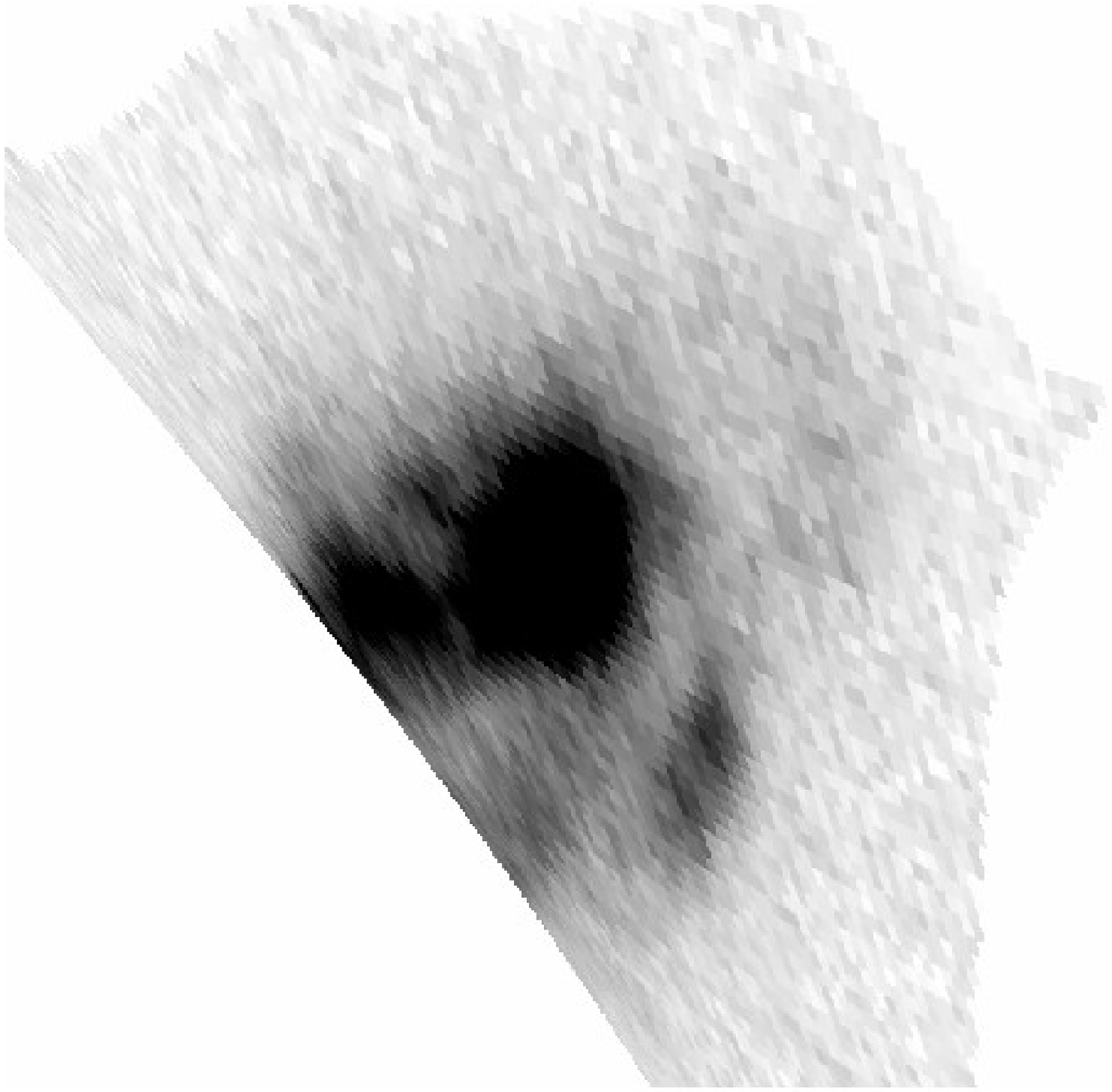} \includegraphics[width=6cm]{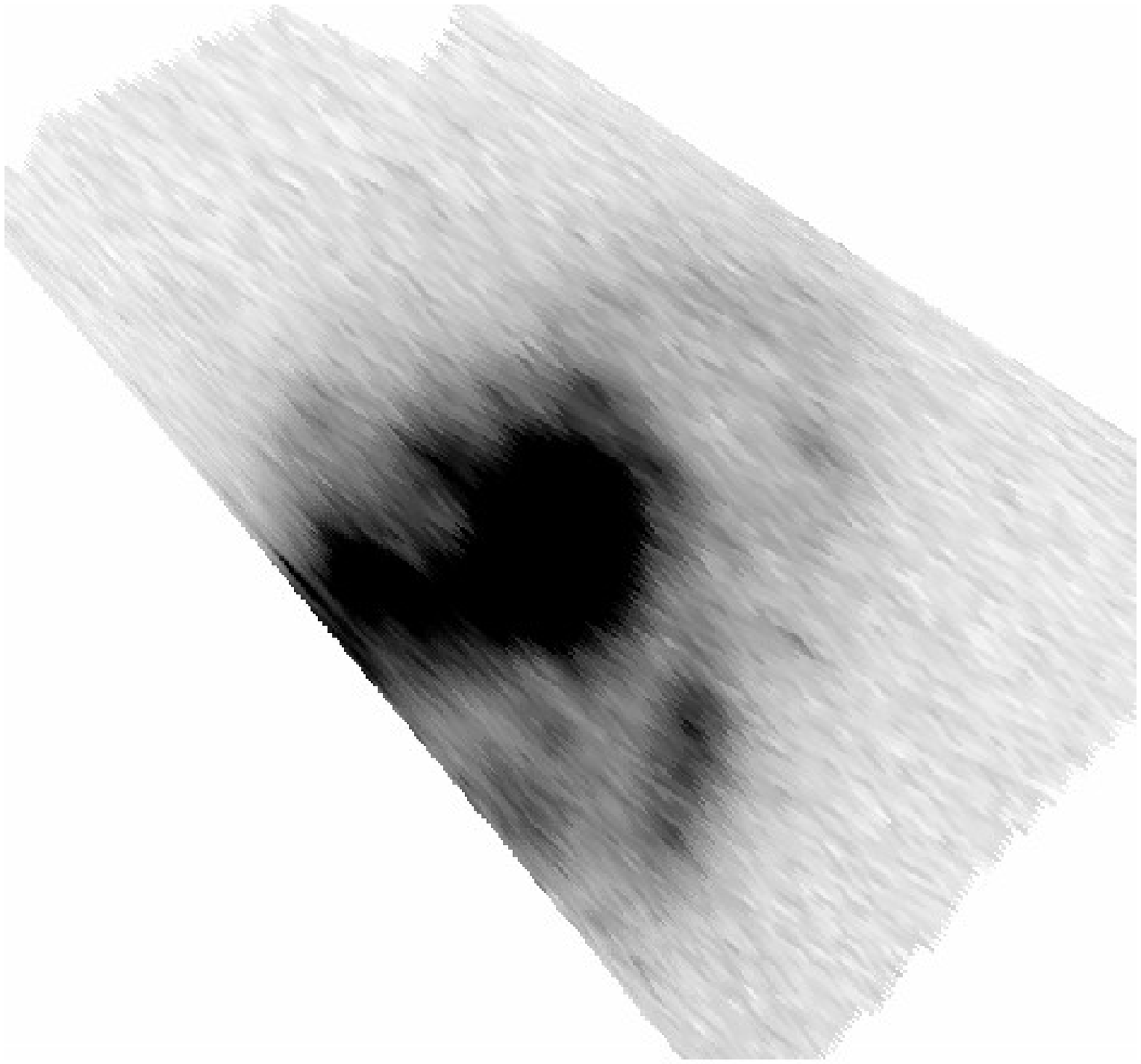}
\includegraphics[width=6cm]{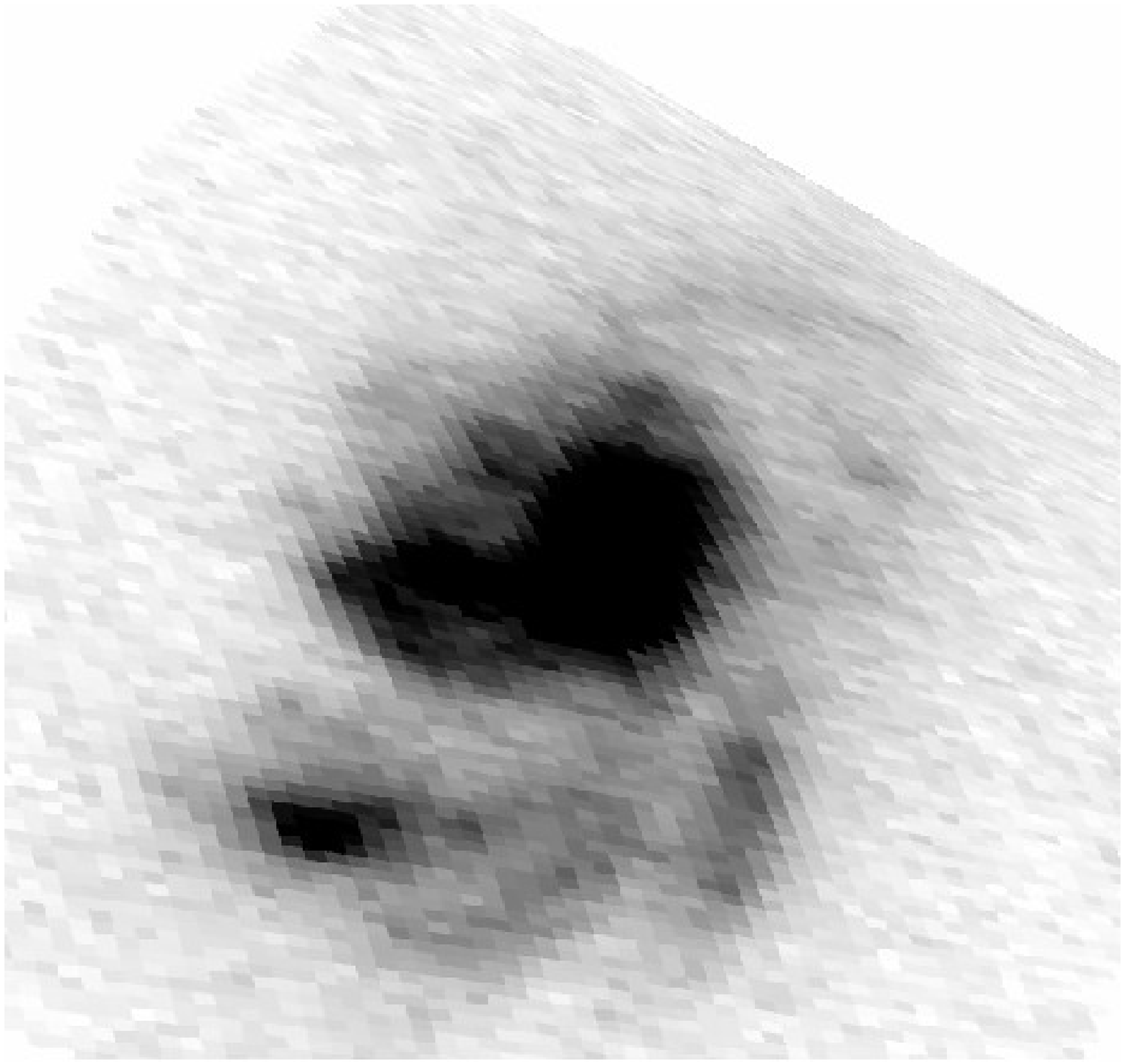} \includegraphics[width=6cm]{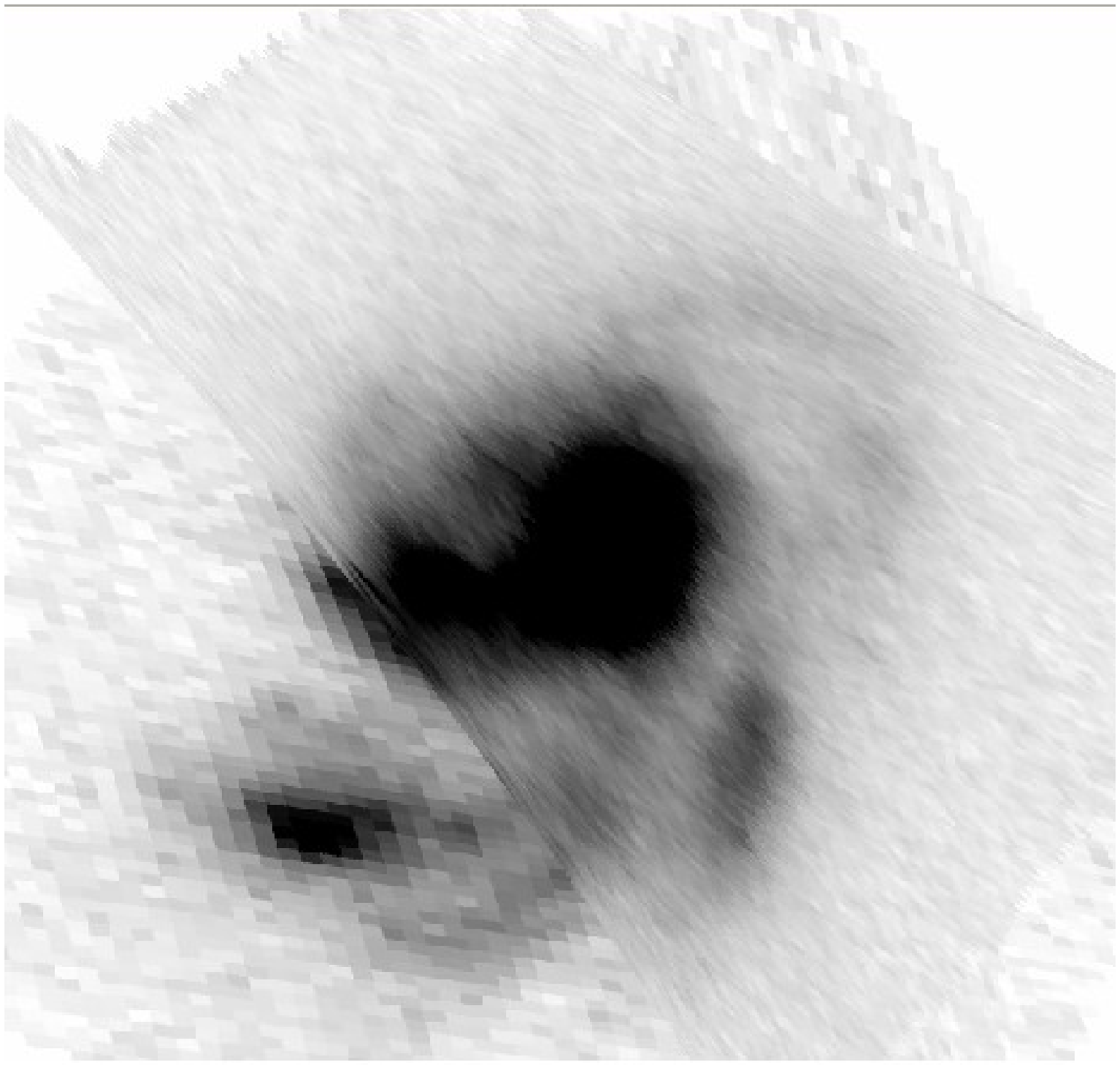}
\caption{ Source reconstruction from each of the magnified images of the giant arc and their rendition into one frame with the best-fit lens model.  \label{fig.sourceplane}}
\end{center}
\end{figure*}

We extract the giant arc within three polygonal apertures
constructed to enclose the whole lensed system (See Fig.~\ref{fig.constraint}).
Fig.~\ref{fig.sourceplane} shows the three independent reconstructions of the
source from the three different lensed images, separately, as
well as the result obtain by considering the three images
simultaneously. The overall agreement between the three
source reconstructions from individual images validates the
best-fit model.

\begin{figure*}
\begin{center}
\includegraphics[width=6cm]{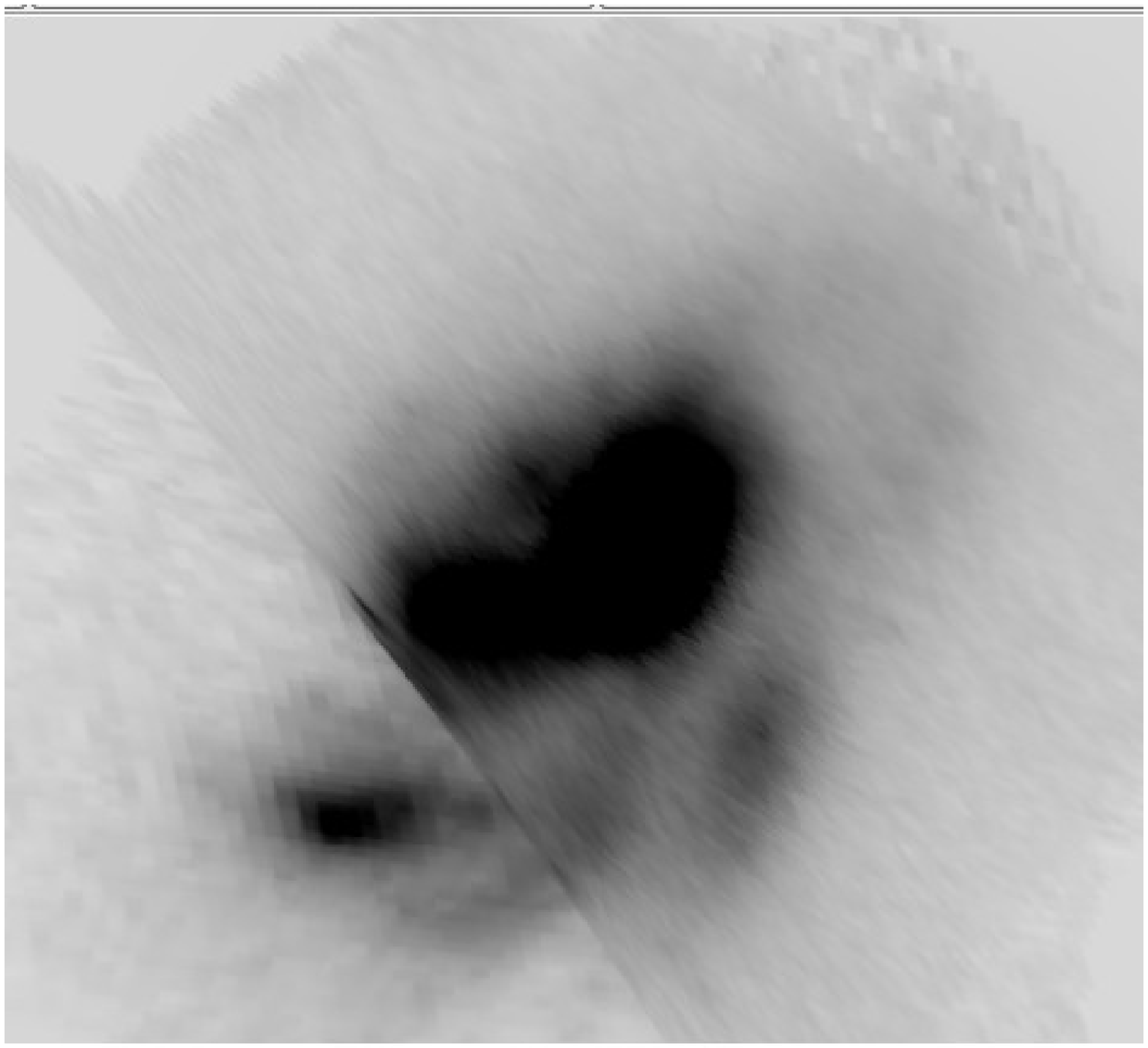} \includegraphics[width=6.3cm]{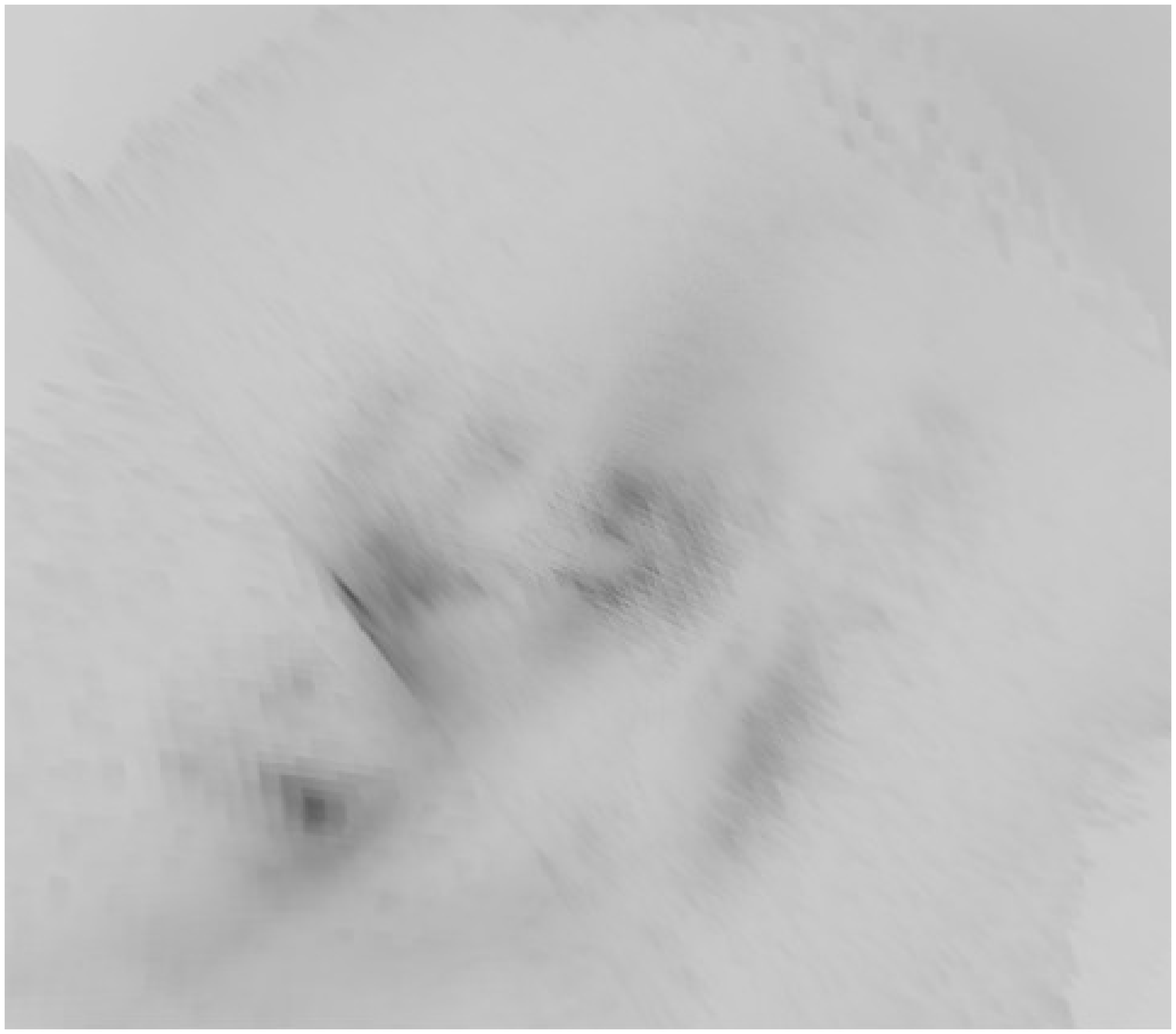}
\caption{Source reconstruction results from MCMC simulations. Left: source reconstruction with the mean value
of lens model parameters; Right: Errors on the source reconstruction given uncertainties on lens model parameters. Note that model-based systematic uncertainties and measurement noise are both included for the reconstructed source-plane image.
\label{fig.MCMC}}
\end{center}
\end{figure*}

In order to take account the uncertainties of the best-fit
lens model parameters, we produce 600 source plane images
through a MCMC simulation, which are then combined to
obtain information on the uncertainties on the reconstructed
source. The simulation results are shown in Fig.~\ref{fig.MCMC}.
The scatter in the pixel values tells us about the impact of the lens model on our reconstruction.
We find the largest systematic errors on the source reconstruction
come from the two clumps on the southeast (about 20\%)
and the faint spiral arms (about 15\%). As for the central bulge
of the galaxy, the effect of reconstruction errors is relatively
small (about 10\%).

On the basis of the reconstructed source from WFPC2
data of the image A3, \citet{Yuan12} could not resolve
between a merging system and a disturbed disk galaxy.
In our reconstruction from the three images, the source
galaxy appears to be an asymmetric disk galaxy with a
bright central bulge, and tightly wrapped spiral arms.
This picture is also confirmed by the regular and smooth
exponential profile shown in the disk (see Section 4.2) and by the
small velocity gradient across the disk found by \citet{Yuan12} based on the observations of the H$\alpha$
emission line. Hence, the source is approximately a face-on rotating disk or a non-rotating system.

\subsection{Multi-band photometry} \label{s.photometry}

A detailed photometric analysis in different wavebands is
required to derive the color profile of the disk galaxy. Therefore,
we used the fiducial best-fit lensing model based on the
HST/ACS $z_{850}$ image (see Table 1) to reconstruct the intrinsic
source morphology in other bands and measure the luminosity
profiles.

\begin{figure*}
\begin{center}
\includegraphics[width=8cm]{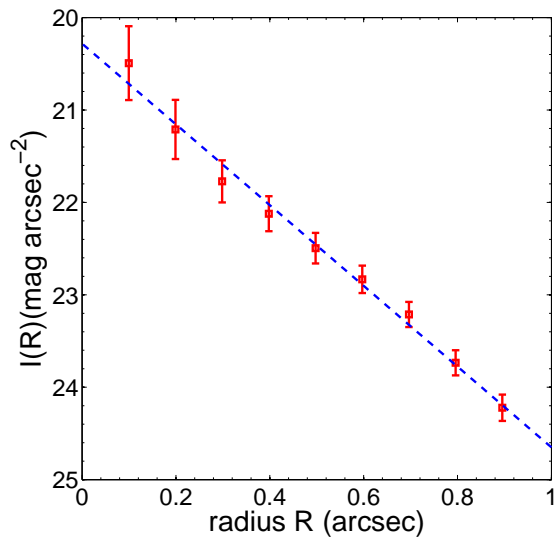}
\caption{Surface brightness of the source galaxy in the $z_{850}$ band. The dashed line is an exponential with $r_s=2.01$ kpc ($h_R=0.25\arcsec$).
Surface brightness is given in units of mag arcsec$^{-2}$: the flux coming from each square arcsecond of the galaxy,
expressed as an apparent magnitude.
\label{fig.brightness}}
\end{center}
\end{figure*}

Fig.~\ref{fig.brightness} shows the surface brightness I(R)(in the unit of mag arcsec$^{-2}$) of the reconstructed
source galaxy from the $z_{850}$-band image. Surface
brightness is measured in elliptical regions, in order to take
into account projection effects: by assuming intrinsic circular
isophotes, we find an inclination angle (along the line of sight) $\theta= 40 \pm 5 $ degrees. In the stellar disk, when averaged over
features like spiral arms, the surface brightness follows approximately
an exponential profile. When expressed in units
of mag arcsec$^{-2}$, this reads
\begin{equation}
I(R)= I(0)+1.086(R / h_R) \,.
\label{IR2}
\end{equation}
At the center of the source galaxy, the $z_{850}$-band surface
brightness is $I(0)=20.28\pm0.22$ mag arcsec$^{-2}$. Fig.~\ref{fig.brightness}
shows the fit to the source \textbf{$z_{850}$}-band surface brightness:
the scale length is $h_R=0.25\pm0.02\arcsec$, corresponding to a
linear radius $r_s=2.01\pm0.16$ kpc at the source redshift.
The apparent magnitude within $r_s$ is $m_s=23.6\pm0.6$.
After correcting for the flux magnification, the intrinsic luminosity
at the galaxy redshift is $L_{*}=2.70^{+0.13}_{-0.08}\times 10^{9}L_{\odot}$.

\citet{Yuan12} have measured the stellar mass of the source galaxy
by fitting the spectral energy distribution (SED). By assuming an extinction
value E(B-V)=0.60 and a star-formation rate SFR$_{SED}=50\pm 35M_{\odot}yr^{-1}$, they estimated the stellar mass to be
log$M_*=10.28\pm0.31$.

We can compare the source of the giant arc in A2667
with the large sample of disk galaxies ta high-z observed
by \citet{Miller11}. Based on spectra from DEIMOS
on the Keck II telescope, \citet{Miller11} measured
the evolving scaling relation of three types of disk-like
galaxies in the redshift range $0.2<z<1.3$. Disk-like galaxies with extended line emission across the 2D spectrum
(like the A2667 giant arc source) have larger mean
scale radii ($r_s=2.68\pm0.09$ kpc) with respect to the compact
($r_s=2.09\pm0.13$~kpc) and passive ($r_s=2.24\pm0.16$~kpc) galaxy samples. Our system has stellar max also
very close to the average stellar mass of disk-like galaxies
in \citet{Miller11}, log$M_*=10.11\pm0.05$. Therefore
the giant arc source appears to be a typical disk-galaxy at $\sim 1$. Note that the passive disk galaxies constitute the upper
end of the total stellar mass with log$M_*=10.69\pm0.09$,
while the typical values are log$M_*=10.21\pm0.10$ for the
compact.

\begin{figure*}
\begin{center}
\includegraphics[width=8cm]{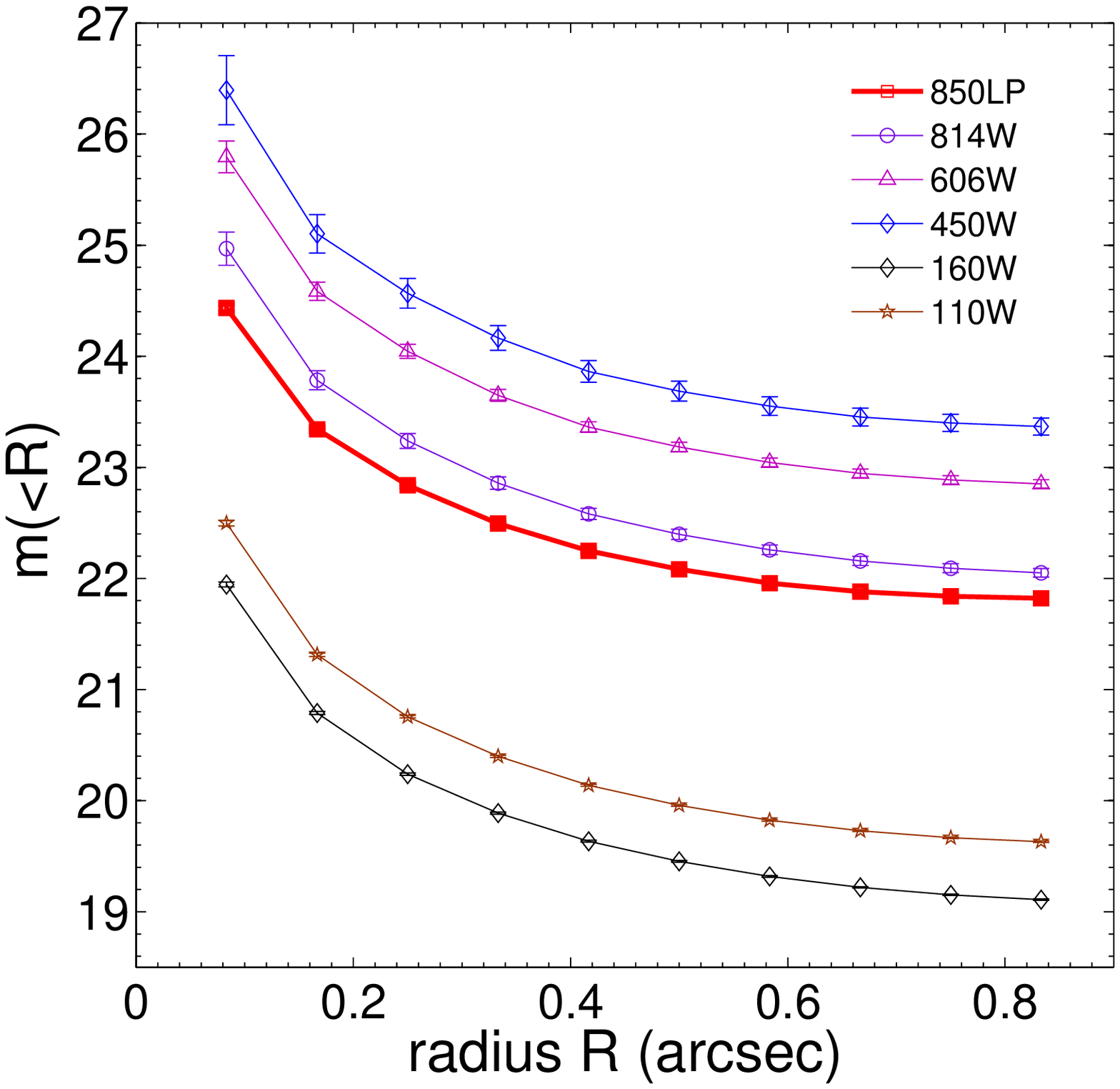}\includegraphics[width=8cm]{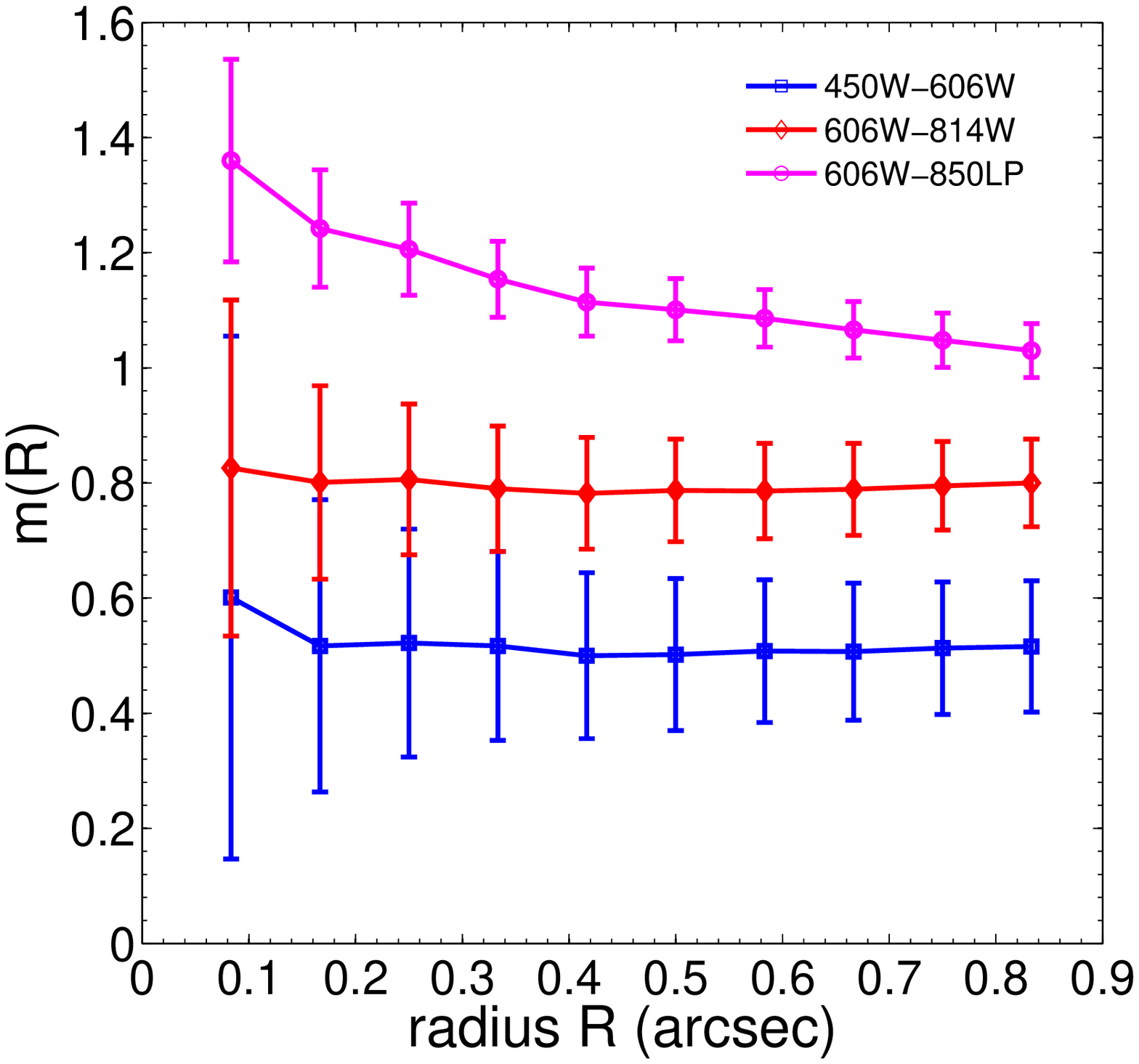}
\caption{The apparent magnitudes and color profiles of the source galaxy at different optical and NIR bands.
\label{fig.gradient}}
\end{center}
\end{figure*}

It is well known that radial color gradient of a spiral
galaxy provides a unique test on the star-formation rate
along the disk and thus the disk formation and evolution.
Multiband photometry of the reconstructed source in six different
bands is presented in Fig.~\ref{fig.gradient}. The lensed galaxy exhibits a
negative gradient (i.e., the color is gradually bluer outwards),
as observed for most of the spiral galaxies \citep{Moth02,MacArthur04,Taylor05}. Moreover,
the color profile becomes shallower with increasing radius
and the color of this source galaxy is redder in the central
bulge, which could be explained by either a standard
bulge/disk structure with the center of this galaxy dominated
by an old stellar population \citep{Yuan12}, or by a merging/
disturbed system that happens to have more attenuation
by dust toward the center of the galaxy.

The above two possible causes can helpfully explain both
the color gradient and the reddened central bulge of the target
galaxy. Regarding the stellar population, it is broadly accepted
that for galactic disks, the star formation rate (SFR) is
proportional to the surface mass density \citep{Bell00,Kauffmann03}, but inversely proportional to the radius \citep{Kennicutt98,Boissier99}. Therefore,
the inner regions of spiral galaxies may possibly evolve faster
than outer regions, which directly leads to older metal-rich
stellar populations in the center \citep{Prantzos00}.
As for the dust, many previous works also showed that its
distribution along the disk of spiral galaxies might be proportional
to the stellar mass density profile  \citep{Regan06},
which gives birth to a inner galactic region containing more
dust and also naturally generates a negative color gradient.

We have measured the color gradient within half-light radius $R_{50}$ and obtained the corresponding values $B_{450}-V_{606}=0.55$ and $V_{606}-I_{814}=0.80$. To make a quantitative comparison, we compare with the previous measurements of
the color difference in galaxies observed in the
GOODS-South Field \citep{Welikala12}, which investigated
the galactic radial variation of colors evolving
with redshift. For the stacked bright sample like our case, the color difference are $0.6<\delta(B-R)<0.8$ for galaxies within the redshift range $0.79<z<0.95$ and $0.25<\delta(B-R)<0.5$ for high redshift galaxies at $1.15<z<1.35$. Concerning the color gradient $\delta(R-I)$ within $R_{50}$, the corresponding values change to $0.8<\delta(R-I)<1.0$ for galaxies for low-redshift galaxies ($0.5<z<1.0$) and $0.5<\delta(R-I)<0.7$ for high-redshift galaxies ($1.0<z<1.5$) \citep{Welikala12}. In our analysis, the calculated color gradients of the lensed galaxy of Abell 2667, a spiral galaxy at $z\sim 1$, are in good quantitative agreement with the previous results above.

\begin{figure*}
\begin{center}
\includegraphics[width=16cm]{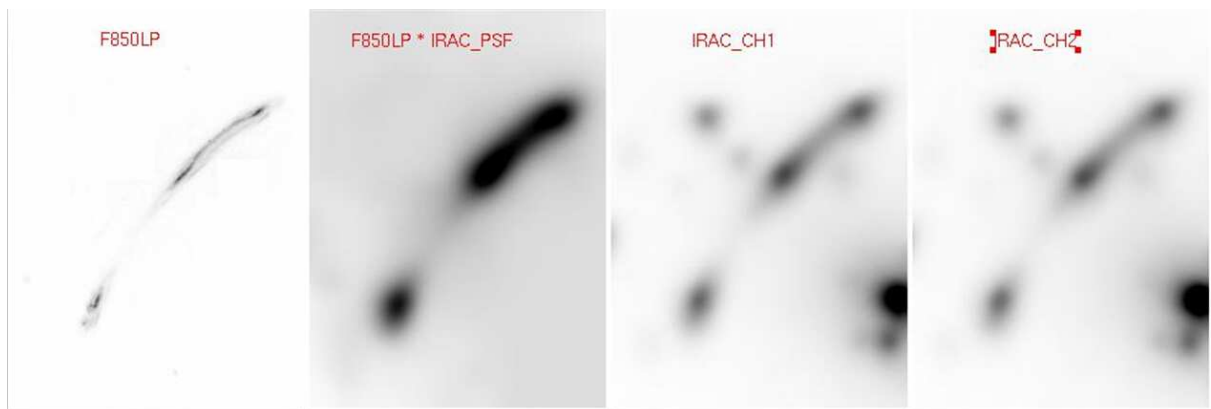}
\caption{The figures show the process used to derive the colors $z_{850}$-mag$_{3.6}$ and $z_{850}$-mag$_{4.5}$ in different regions of the gravitational arc.
From the left, the panels show: the image of the arc in the HST/ACS, the HST/ACS data convolved with the Spitzer/IRAC PSF,
the arc observed with Spitzer/IRAC at 3.8 and 4.5 $\mu {\rm m}$.
\label{fig.color}}
\end{center}
\end{figure*}

In order to determine the nature of the reddened galactic center, firstly, we make measurements of the apparent magnitude of the galaxy center within $0.17\arcsec$ (which corresponds to 1.34 kpc). As can be seen in Fig.~\ref{fig.gradient}, the center of this galaxy is much redder (B$_{450}$-V$_{606}$=0.50). Secondly, the colors $z_{850}$-mag$_{3.6}$ and $z_{850}$-mag$_{4.5}$ are also measured in the image plane. In order to perform photometric measurement on the Spitzer IRAC data, we convolve the $z_{850}$ HST image with the IRAC PSF and subtract it, after rescaling, from the IRAC images (see Fig.~\ref{fig.color}). As expected, the colors of the arc are quite different between the center of the spiral and the spiral arms. We find that in the bulge $z_{850}$-mag$_{3.6}$=0.23 and $z_{850}$-mag$_{4.5}$=0.03. In the spiral arms, the color is much bluer: $z_{850}$-mag$_{3.6}$=-0.36 and $z_{850}$-mag$_{4.5}$=-0.72. The uncertainty on these colors is $\sim 0.1$mag, as evaluated from the amount of residuals in the respective regions.

Theory of dust attenuation
has been reviewed many times in different papers \citep{MacArthur04,Taylor05}. However,
we lack theories with dust attenuation only to explain
the fact that the center of this galaxy is too much
brighter in the R band than the B band. From the observational
point of view, galaxies with high surface mass
density always have more dust and generate steeper color
gradients \citep{Regan06,Liu09}, i.e. the
more luminous galaxies have steeper scaled color gradients,
which is inconsistent with our observations. Consequently,
dust attenuation only plays a minor role in regulating
the color gradients of spiral galaxies, or at least
could not be the only role in the origin of reddened galactic
center \citep{MacArthur04,Taylor05}.

Thus, we can reach a careful conclusion that the central
bulge of the lensed galaxy of Abell 2667 tends to be heavily
dominated by an old stellar population, meanwhile, the other
two blue clumps on the south are very likely to be active starforming
regions and responsible for the starbursts of this blue
star-forming galaxy \citep{Yuan12}. Further constraints
and discussion on the spectroscopic type of this star-forming
galaxy are given in Section~\ref{s.spectroscopy}.

\subsection{VIMOUS/IFU spectroscopy} \label{s.spectroscopy}

\begin{figure*}
\begin{center}
\includegraphics[width=9cm, angle=0]{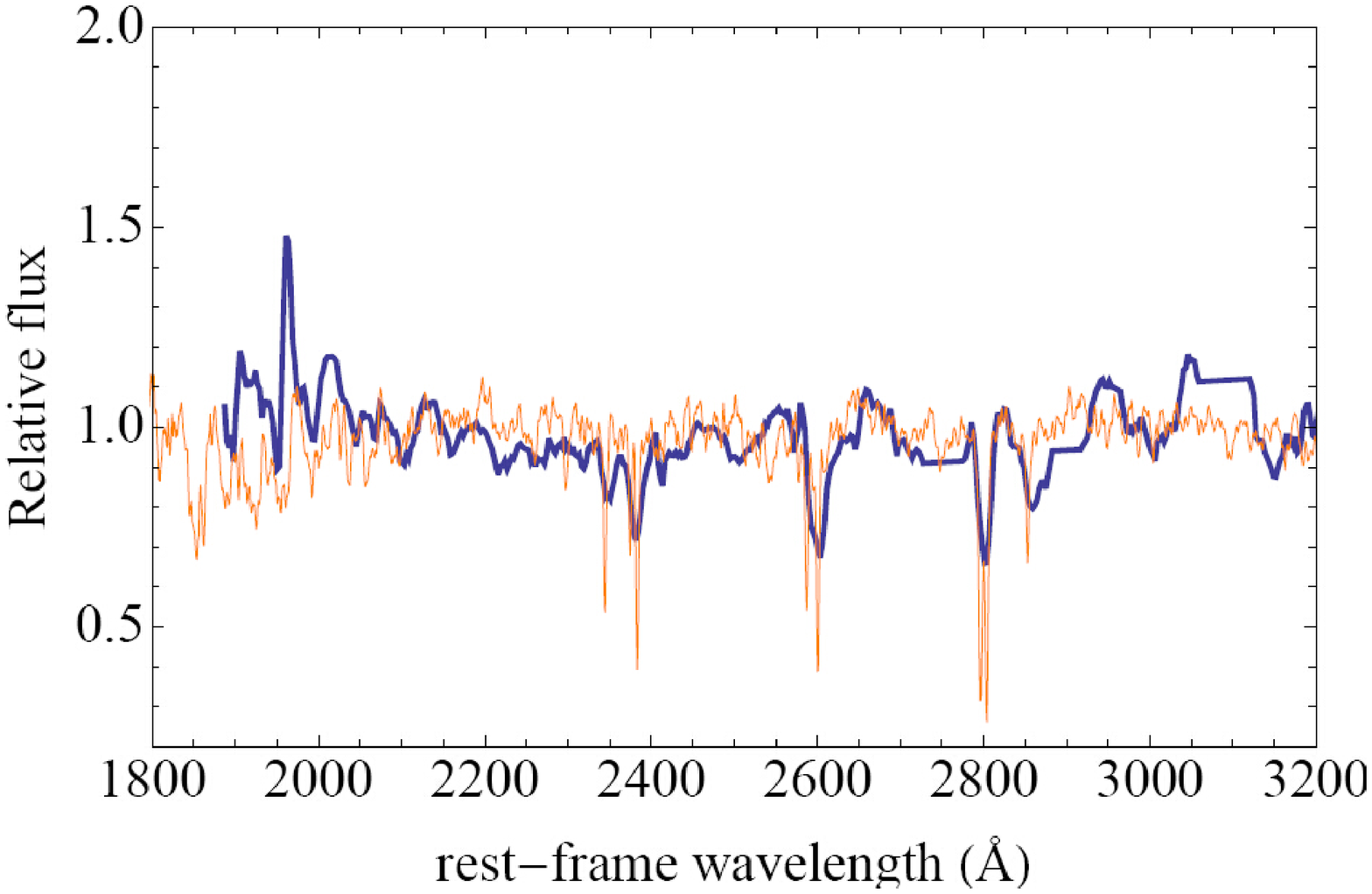}\includegraphics[width=9cm, angle=0]{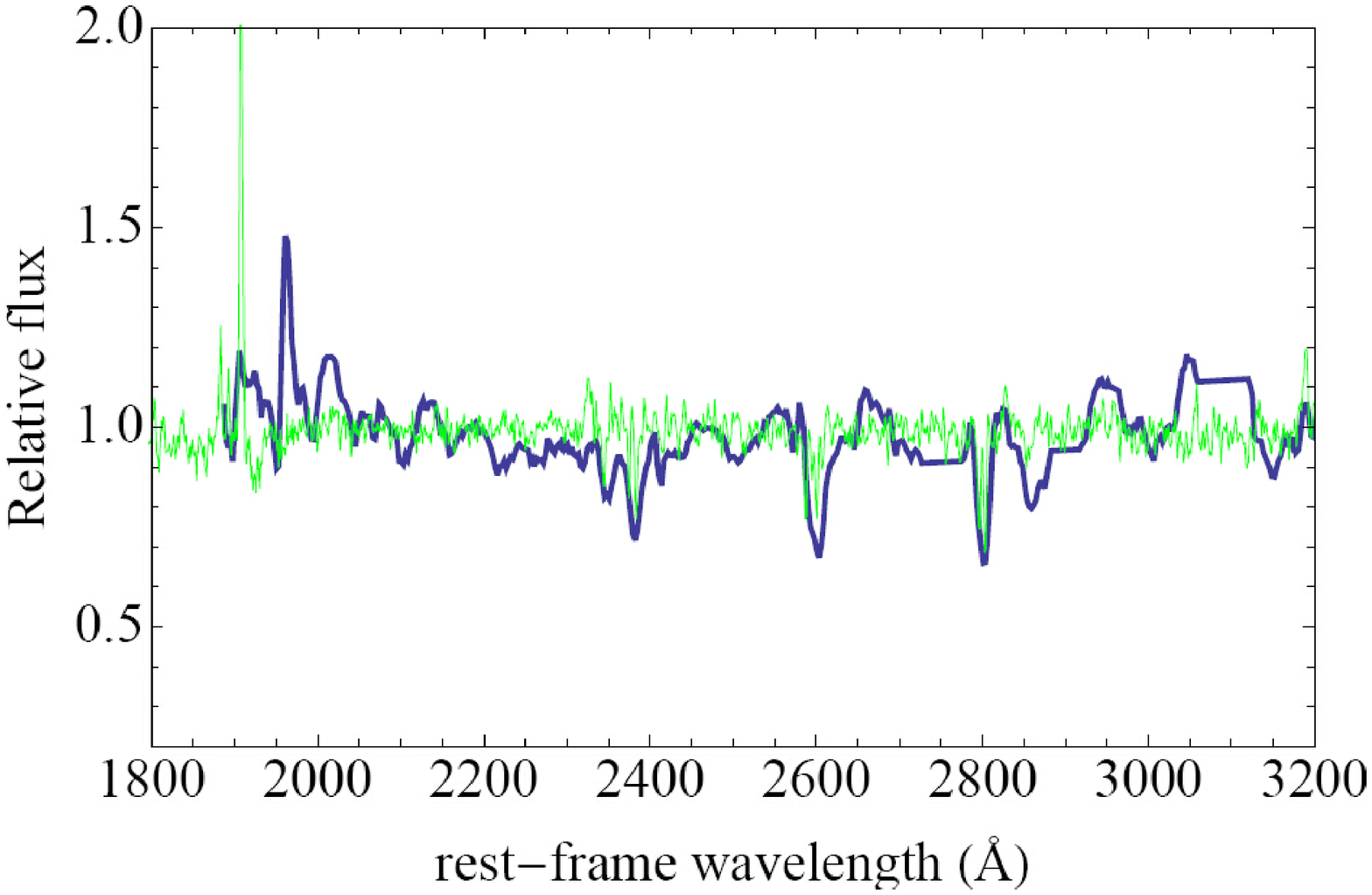}
\includegraphics[width=10cm, angle=0]{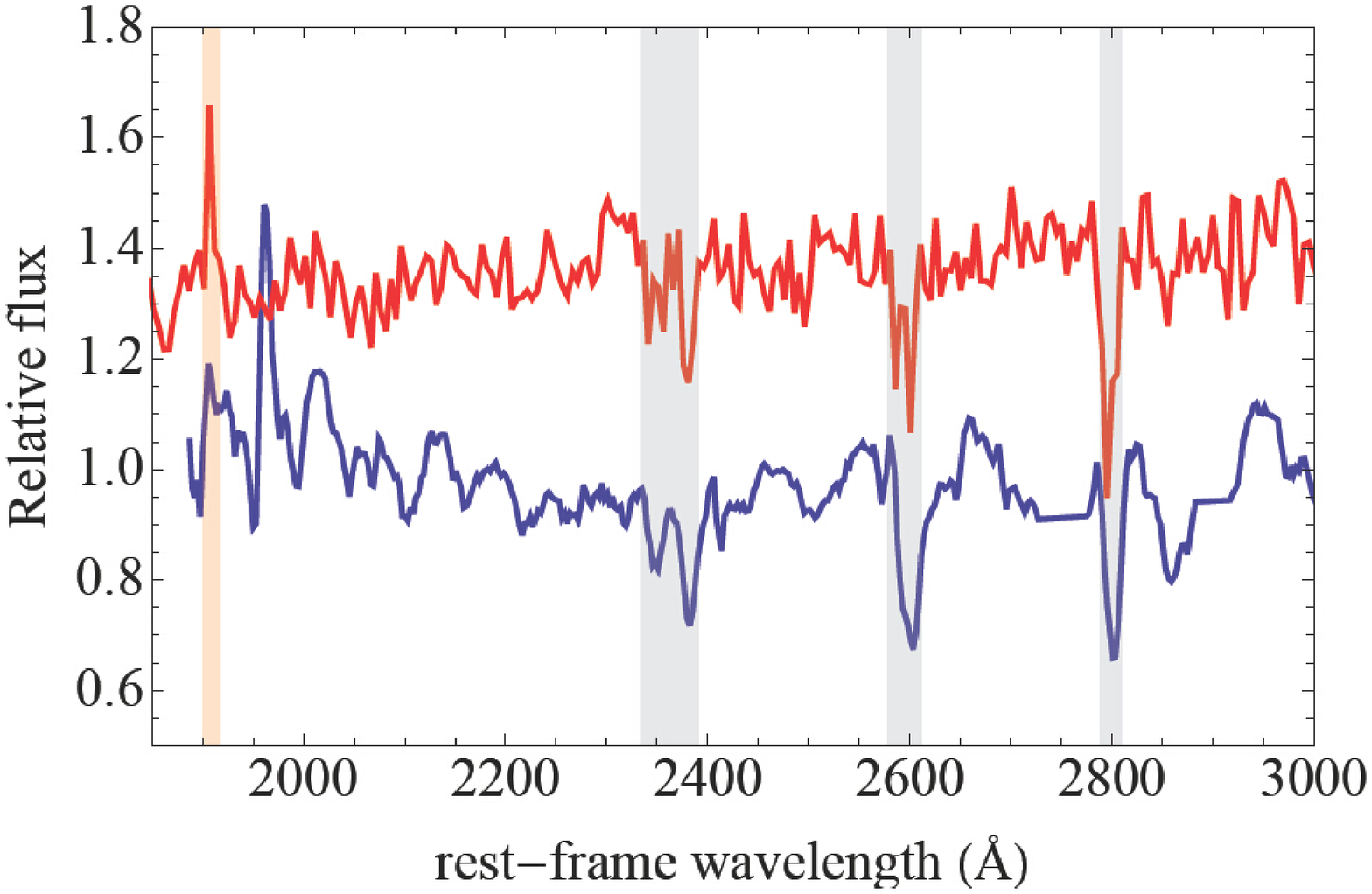}
\caption{The arc spectrum of Abell 2667 (blue spectrum) compared with two templates of high (orange spectrum) and low-metallicity (green spectrum) starburst galaxies, as well as the UV spectrum of the local Wolf-Rayet galaxy NGC~5253 (red spectrum). The prominent features (e.g., CIII] emission and some of the strong wind absorption features) that motivate the classification as a Wolf-Rayet galaxy spectrum are also identified. The gray shaded lines denotes the EWs of FeII and MgII in the Abell 2667 spectrum considering the same wavelength interval of the NGC 5253 spectrum, while the orange shaded line marks the presence of the CIII] line.
\label{fig:spec}}
\end{center}
\end{figure*}

We discuss here the VIMOUS/IFU spectroscopic data collected by \citet{Covone06b}, in the given wavelength range ($\sim 1800-3200$ \AA). Fig.~\ref{fig:spec} shows the combined spectrum in blue line from the three images of the the giant gravitational
arc (i.e., summing the signal from all the three
images of the arc). The source shows the characteristic
spectrum of a blue star-forming galaxy, with a strong UV
continuum emission and evident spectral features due to
low-ionization absorption lines (In particular, the strongest in our spectra are the CIII] 1909 line, the FeII 2370/2600 lines, and the MgII/2800 line).

We have compared the spectrum of the arc on Abell 2667 with some galaxy templates, focusing particularly on starburst galaxies.
Following the work of \citet{Leitherer11}, we have first compared the spectrum of the arc on Abell 2667 with two templates of high (orange spectrum) and low-metallicity (green spectrum) starbursting galaxies in Fig.~\ref{fig:spec}. The high-z spectrum is characterized by an almost absence of the CIII] 1909 line and more pronounced interstellar lines of FeII, MnII
and MgII, while in the low-z spectrum the CIII] nebular line is very pronounced, but a lower intensity for interstellar lines. In the spectrum of Abell 2667 these absorption lines are all present, including the MgI 2853 line. The resolution of the spectrum does not allow to split the single FeII-MnII line and the MgII doublet at 2796-2803 \AA.

The presence of the CIII] line suggests a further spectra comparison with the starbursting Wolf-Rayet galaxy NGC5253, from the database complied by \citet{Storchi04,Storchi05}.
Wolf-Rayet (WR) galaxies are a subset of blue emission-line
galaxies showing the signature of large numbers
of WR stars in their spectra \citep{Vacca92}. WR galaxies are thought
undergoing present or very recent star formation, with a large
population of massive stars in the age range $1 \sim 10 \times 10^6$ yr.
NGC 5253 is a blue compact WR galaxy at distance $3.6\pm0.2$ Mpc \citep{Sakai04},
showing an intense burst of star formation in the central region.
With the measurements of the equivalent widths of blended interstellar lines for NGC 5253 \citep{Leitherer11},
we have measured the EWs in the arc on Abell 2667 spectrum considering the same wavelength interval of the NGC 5253 spectrum (gray shaded lines in Fig.~\ref{fig:spec}).

We obtain EW(FeII 2370)= 5.42 $\pm$ 0.58, EW(FeII 2600) = 6.77 $\pm$ 1.32, and EW(MgII 2800)=5.66 $\pm$ 1.26, similar to the values obtained for the NGC5253 spectrum: EW(FeII 2370) = $5.40\pm0.82$ , EW(FeII 2600) = $4.56\pm0.51$, and EW(MgII 2800) = $5.27\pm0.35$. We then first infer a quite low-metallicity (log[O/H] $\leq 8.5$) for the lensed galaxy Abell 2667.
In Fig.~\ref{fig:spec} it is also shown the presence of the CIII] line in the spectrum of NGC 5253 with the one in the arc on Abell 2667 spectrum (orange shaded line). The low-significance CIII] line in the spectra is likely a product of the low resolution VIMOS spectra, which makes it difficult to identify lines with relatively small equivalent widths \citep{Bayliss13}.

%Moreover, we find that, in the observed spectral range of the lensed galaxy of Abell 2667, the value of UV slope $\beta$, defined by the relation %$F_\lambda\varpropto \lambda^\beta$,
%is close to that of the star-forming galaxies ($\beta=-1.11\pm 0.44$) in the redshift range $1< z < 2$ \citep{Talia12}.

Meanwhile, we remark that the physical origin of CIII] emission is not entirely certain, as it should in general be offset in
velocity from absorption features that are result from the
large gas outflows that WR stars create. To fully understand
the classification of the lensed galaxy, a higher resolution
spectrum is required to measure the velocity offsets between the CIII] emission and the FeII and MgII absorption
lines.

\section{Discussion and conclusions}\label{s.conclusions}

We used HST/ACS z$_{850}$ imaging data to construct an improved
lens model for the lensing cluster Abell 2667. Using
the new lens model, we reconstructed the de-lensed image of
the source galaxy at $z\sim 1$. The source resembles a normal
disk galaxy with a bright, large central bulge, and tightly
wrapped spiral arms. Multiband imaging data allowed us
to identify features of the lensed source galaxy in different
bands. We summarize our main findings here:

1. The surface brightness $I(R)$ (Fig.~\ref{fig.brightness}) of the reconstructed source galaxy from the CCD image in the z$_{850}$ band reveals
the central surface brightness $I(0)=20.28\pm0.22$ mag arcsec$^{-2}$ and the exponential slope with scale length $h_R=0.25\pm0.02\arcsec$ (a characteristic radius $r_s=2.01\pm0.16$ kpc at redshift $z=1.0334$).  The smooth exponential profile supports the
hypothesis of mildly disturbed, almost face-on, disk versus
a merging system, consistent with the velocity structure observed
by \citet{Yuan12}.

2. After correcting for the flux magnification, the intrinsic
luminosity of this spiral galaxy is $L_{*}=2.70^{+0.13}_{-0.08}\times 10^{9}L_{\odot}$,
with the corresponding stellar mass log$M_*=10.28\pm0.31$.
These values are close to the ones of disk galaxies with extended
emission lines observed at $z>1$ \citep{Miller11}.

3. There is negative radial color gradients along the disk
(i.e., the color is gradually bluer outwards). The color profile
becomes shallower with increasing radius, which provides
unique star-forming information along the disk and thus the
history of disk formation and evolution. Moreover, we find
that the central region of the galaxy tends to contain more
metal-rich stellar populations, rather than heavily reddened
by dust due to high and patchy obscuration.

4. We further analyze the archive VIMOUS/IFU spectroscopic
data \citep{Covone06b} and find that the spectra of
the source galaxy with the giant gravitational arc is characterized
by a strong continuum emission with strong UV absorption
lines (FeII and MgII). Moreover, in the given wavelength
range ($\sim 1800-3200$ \AA) of the combined arc spectrum, the
lensed galaxy of Abell 2667 shows some typical features of a typical
starburst WR galaxy (NGC 5253) with strong signatures from
large numbers of WR stars (Fig.~\ref{fig:spec}). More specifically, we have measured the EWs in the Abell 2667 spectrum considering the same wavelength interval of the NGC5253 spectrum, and obtained EW(FeII 2370)= 5.42 $\pm$ 0.58, EW(FeII 2600) = 6.77 $\pm$ 1.32, and EW(MgII 2800)=5.66 $\pm$ 1.26, which are similar to the values for the NGC5253 spectrum. Marginal evidence for CIII] 1909 emission at the edge of the grism range, as expected for a typical WR galaxy, further confirm our expectation. However, this conclusion still needs to be checked by high resolution spectrum covering different wavelength range.

\section*{Acknowledgments}
We thank the anonymous referee for comments which improved the presentation of the paper. This work was supported by the Ministry of Science and Technology National Basic Science Program (Project 973) under Grants Nos. 2012CB821804 and 2014CB845806,
the National Natural Science Foundation of China under Grants Nos. 11073005 and 11373014,
the Fundamental Research Funds for the Central Universities and Scientific Research Foundation of Beijing Normal University,
and China Postdoctoral Science Foundation under Grant No. 2014M550642. We acknowledge financial support from the grant PRIN-INAF 2011 "Galaxy
Evolution with the VLT Survey Telescope".

\end{document}